# Multiple Scattering of Fractionally-Charged Quasiparticles

E. Comforti, Y. C. Chung, M. Heiblum, and V. Umansky

Braun Center for Submicron Research, Department of Condensed Matter Physics, Weizmann Institute of Science, Rehovot
76100, Israel

(November 21, 2018)

We employ shot noise measurements to characterize the effective charge of quasiparticles, at filling factor $\nu = 1/3$ of the fractional quantum Hall regime, as they scatter from an array of identical weak backscatterers. Upon scattering, quasiparticles are known to bunch, e.g., only three $e/3$ charges, or 'electrons' are found to traverse a rather opaque potential barrier. We find here that the effective charge scattered by an array of scatterers is determined by the scattering strength of an individual scatterer and not by the combined scattering strength of the array, which can be very small. Moreover, we also rule out intra-edge equilibration of $e/3$ quasiparticles over length scale of hundreds microns.

Shot noise measurements, being sensitive to the charge of carriers, have recently become a major tool for characterizing the quasiparticle charge in the fractional quantum Hall (FQH) regime. Measurements are performed by introducing a weak potential scatterer in the path of an otherwise noiseless current, thus leading to stochastic partitioning of the quasiparticles and to shot noise proportional to their charge. The existence of quasiparticles with charge $e^* = e/3$ and $e^* = e/5$, in the $\nu = 1/3$ [1,2] and $\nu = 2/5$ [3] FQH states, respectively, was verified. Quasiparticles in edge states are thought to be independent since their shot noise agrees well with an independent-particle model. This model, however, is applicable only in the case of *weak backscattering*, namely, when the forward transmission $t$ of the scattering potential approaches unity. At $\nu = 1/3$, as $t$ decreases, correlation among the transmitted quasiparticles is being induced by the scattering potential. In the strong backscattering limit ($t \to 0$), three quasiparticles must group to an 'electron' in order to tunnel through the barrier [4]. The induced correlations among quasiparticles are parameterized by effective (heuristic) independent particles with charge $q$, each made of independent *bunched quasiparticles* [5]. When $q$ is measured over the whole range of transmission ($t \in [0, 1]$), it evolved smoothly from $q = e/3$ at $t \approx 1$ to $q \approx e$ at $t \approx 0$ - indicating an enhanced correlation as the transmission diminishes.

Systems containing a single backscatterer are easy to analyze since only the transmission $t$ determines the charge. On the other hand, a cascade of $N$ identical potential scatterers ($t_j \equiv t \leq 1$ for all $j$) requires a different analysis. In the absence of multiple reflections between the scatterers the total transmission, given by $t_0 = t^N$, can be made arbitrarily small for large enough $N$. A 'macroscopic' model, based on a single *effective barrier* with transmission $t_0$, leads to an effective charge determined by such barrier. One expects in the case of $t_0 \to 0$ the effective charge to be $e$. Contrarily, a 'microscopic' model of an array of weakly backscattering potentials points out the absence of a high potential barrier (believed to be the origin of quasiparticle bunching), hence leading to an effective charge corresponding to a single barrier in the chain, e.g., $q(t \to 1) = e/3$ (independent on the $t_0$). We measure here $q$ in such system.

When analyzing shot noise produced by multiple scattering we use the superposition theorem [6], namely, we sum the contributions to the noise of individual scatterers. Consider, for simplicity, the case of two sequential scatterers of *electrons*, characterized by the same transmission $t$. The first scatterer, impinged by a noiseless incident current $I_{inc}$, produces, at zero temperature, noise $2eI_{inc}t(1-t)$ [7]. This noise is further attenuated by $t^2$ when traverses the second scatterer, resulting with contribution to the total noise given by

$$S_1 = 2eI_{inc}t(1-t) \cdot t^2 \ . \qquad (1)$$

We add now the noise *generated* by the second scatterer. This scatterer is impinged by a current $I_{inc}t$, which according to the superposition theorem we assume it is noiseless (the impinging noise was already taken into account in Eq. (1)). It therefore generates noise given by

$$S_2 = 2e(I_{inc}t)t(1-t) \ . \qquad (2)$$

One gets for the total noise the expected expression:

$$S = S_1 + S_2 = 2eI_{inc}t^2(1-t^2) = 2eI_{inc}t_0(1-t_0) \ , \qquad (3)$$

with $t_0 = t^2$ the total transmission. The generalization to a cascade of $N$ scatterers is similar with $t_0 = t^N$. An expression similar to Eq. (3) is used when analyzing the scattering of *quasiparticles* having transmission-dependent effective charge $q(t)$, with $q(t)$ replacing $e$ in the individual noise expressions (Eqs. (1) and (2)). However, the suppression factor $1-t$ must be replaced by $1-\tilde{t}$, where $\tilde{t} = t\frac{e/3}{q(t)}$ is the charge-dependent transmission of particle flux, in order to preserve the conductance of the system, $G \propto t \cdot q(t)$ (the first $qt$ term stands for the actual conductance, hence should not be replaced by $\tilde{qt}$). Hence, the total noise can be approximated as

$$S \approx 2q_0(t)I_{inc}t_0(1-\tilde{t_0}) \ . \qquad (4)$$

The expression is exact for either $t \approx 1$ (with $q = e/3$ and $\tilde{t} = t$) or $t \ll 1$ (with $\tilde{t} \ll 1$) [8]. We refer to Eq. (4) as the 'microscopic' model, since it contains the *local* charge $q(t)$ as the ratio between total noise and transmitted current.

If physics is different and the system behaves as a single effective scatterer with transmission $t_0$, one can assign to it an effective charge $q_0(t_0)$, and the equivalent of Eq. (4) is

$$S = 2q_0(t_0)I_{inc}t_0(1-\tilde{t_0}) \ . \qquad (5)$$

We thus refer to Eq. (5) as the 'macroscopic' model. Fitting the experimental data to Eq. (4) or (5) enables discrimination between the two models.

Samples are fabricated in a high mobility, low density, two-dimensional electron gas (2DEG), with mobility $2 \times 10^6 \text{cm}^2/\text{Vs}$ and areal carrier density $1.1 \times 10^{11} \text{cm}^{-2}$, both measured at 4.2 K in the dark. FQH state, with bulk filling factor $\nu = 1/3$, is achieved by applying a magnetic field $B \approx 13$ T. Vanishing of the longitudinal resistivity $\rho_{xx}$ assures that the (net) current flows chiraly along the edges of the sample as *edge states*. This allows measurements in a multi-terminal geometry described schematically in Fig. 1(a). The actual realization, shown in Figs. 1(c) and 1(d), has metal gates (unlabeled), that form barriers in the 2DEG when negatively biased, and Ohmic contacts to the 2DEG (labeled). The standard geometry, shown schematically in Fig. 1(a), is made of a cascade of quantum point contacts (QPCs), each partitions the current by the same ratio: $t$ transmitted and $r$ reflected. Noiseless dc current, $I_{inc}$, is injected at $S$, partly transmitted toward the next QPC and partly reflected to a grounded drain $D_1$. This process repeats itself $N$ times, with total transmission $t_0 = \prod_{j=1}^N t_j$, thus diluting the current with increased $N$. The grounded drains between consecutive QPCs collect the reflected currents and prevent multiple reflections. The noise at $A$ is first amplified by a cooled homemade amplifier and then measured by a spectrum analyzer. The cooled amplifier has low current-noise at its input, $\langle i^2_{amp} \rangle = 1.5 \times 10^{-28} \text{A}^2/\text{Hz}$, when it operates around a center frequency $f_0 \approx 1.5$ MHz (chosen to be well above the cutoff of the ubiquitous $1/f$ noise). $f_0$ is determined by resonance of a $LC$ circuit (bandwidth 30 kHz), with $C$ dominated by the unavoidable capacitance of the coaxial cable connecting the sample and the amplifier and $L$ set by an added superconducting coil (see Ref. 1 for details). Note that the injected and reflected currents do not share the same terminals: in the input they are source $S$ and grounded $D$, and at the output they are output $A$ and grounded $T$. Hence, both the *input* and *output* impedances are constant, $G = \frac{e^2}{3h}$ at $\nu = 1/3$, and independent of QPCs' transmissions. This leads both the sample's equilibrium thermal noise $(4k_BTG)$ and the back-fed noise of the amplifier $(\langle i^2_{amp}\rangle/G^2)$ to be independent of all $t_j$'s, allowing subtracting them from the total noise [9].

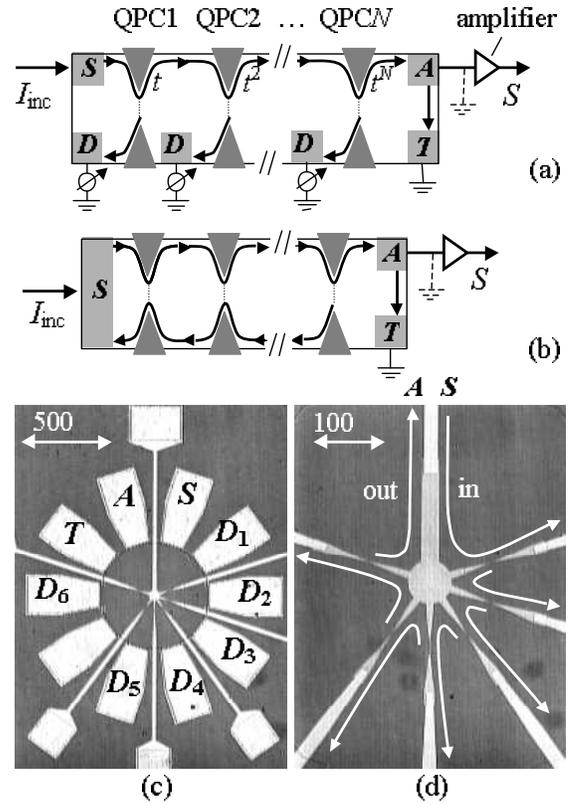

FIG. 1. **(a)** Multiple scattering of quasiparticle current in a *multi-terminal* geometry: a dc (noiseless) incident current from $S$ is partitioned when transmitting via a cascade of QPCs, with the resulting noise measured by a cooled, low-noise, amplifier at $A$. The intermediate drains $D$ prohibit multiple reflections and enable determination of the individual transmission of each QPC. The grounded terminal $T$ is used to fix the output impedance. **(b)** Removing the intermediate drains results with a *two-terminal* geometry. **(c)** Photograph of an actual device, designed according to geometry (a). Labeled elements are Ohmic contacts, unlabeled are gates. **(d)** Close-up of the vicinity of the QPCs, with the direction of the current along the edges.

The experimental data is actually fitted to a modified expression of shot noise that accounts for the finite temperature of the quasiparticles (65 mK, as extracted from thermal noise measurements) [10]:

$$S = \sum_{j=1}^N 2q_jI_j(1-\tilde{t})\left[\coth\left(\frac{q_jV_j}{2k_BT}\right) - \frac{2k_BT}{q_jV_j}\right] \cdot t^{2(N-j)} \ ,$$
(6)

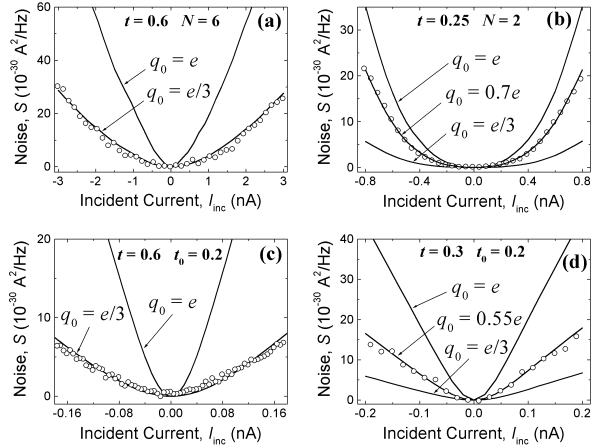
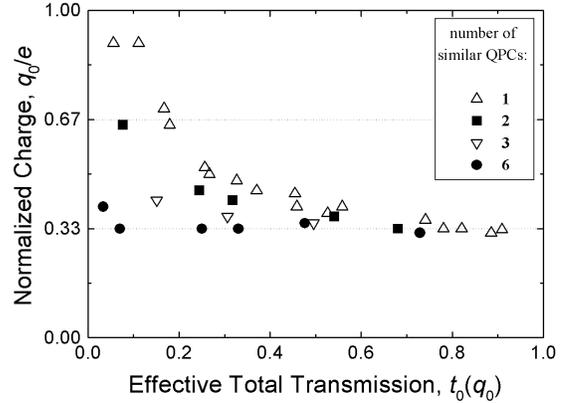

FIG. 2. Noise measurements in various multiple-scattering configurations. **(a)** Six weakly backscattering QPCs ($t = 0.6$) in the multi-terminal geometry. The total transmission is $t_0 = t^6 = 0.05$. **(b)** Similar total transmission achieved by cascading two moderately backscattering QPCs ($t = 0.25$, $t_0 = t^2 = 0.06$) in the multi-terminal geometry. **(c)** Five weakly backscattering QPCs in the two-terminal geometry. The total transmission is measured to be $t_0 = 0.2 > 0.6^5$ due to multiple reflections between the QPCs. **(d)** Two moderately backscattering QPCs ($t = 0.3$) in the two-terminal geometry, resulting with $t_0 = 0.2$ as well. In all cases the total charge $q_0$ coincides with the local charge $q$.

where $I_j$ and $q_j V_j$ denote the current transmitted through, and the highest energy of the impinging particles, at the $j^{\text{th}}$ QPC, respectively. The term in the square brackets, $X_j(q_j V_j, T)$ for short, is responsible for the smooth crossover from the thermal (equilibrium) noise at $qV \ll k_B T$ to the asymptotic shot noise (linear with current) at $qV \gg k_B T$. Note also that the nonlinearity of $t$ is taken into account by integrating over energy [5]. We plot in Fig. 2(a) the shot noise generated by six similar QPCs with total transmission $t_0 = 0.05$ (while $t = 0.6$). Each QPC is found to partition quasiparticles with an effective charge $q(t = 0.6) \approx e/3$ when is impinged by an undilute current. As seen, the noise of the total system corresponds closely to an effective charge $q_0 = e/3$ - the same as that of a single QPC, and not to $q_0(t_0 = 0.05) \approx e$. In comparison, the measured noise of only two QPCs with $t_0 = 0.06$, each having $t = 0.25$ and $q(t = 0.25) = 0.7e$, is shown in Fig. 2(b). We find $q_0 = 0.7e$ - agreeing again with the 'microscopic' and not the 'macroscopic' model. In other words, quasiparticles are bunched by local 'potential bottlenecks' (a pinched QPC) but remains $e/3$ as long as the $\nu = 1/3$ state is only weakly perturbed.

To tie these results more directly with single barrier transport we remove the intermediate drains converting thus the configuration to a two-terminal one (shown in Fig. 1(b)). Consequently, when $t_0$ is made small enough, the sample is *completely* divided between **S** and **A** by an

FIG. 3. Total charge *vs.* total (effective) transmission for different number of QPCs $N$. When $N$ is large, a dilute beam (total transmission $t_0 \ll 1$) of individual quasiparticles (with charge $q_0 = e/3$) can be produced.

*effective insulator* [4]. The total transmission $t_0$ is found to be larger than $t^N$ (due to multiple reflections between QPCs). For example, for $t = 0.6$ and $N = 5$, we find $t_0 = 0.2 > 0.6^5$, and for $t = 0.3$ and $N = 2$, we find $t_0 = 0.2 > 0.3^2$. In these examples, the measured noise shown in Figs. 2(c) and 2(d) corresponds to $q(t)$ rather than to $q_0(t_0)$ - verifying again the 'microscopic' model.

We summarize in Fig. 3 the effective charge $q_0$, gathered by various measurements in a multi-terminal configuration, as function of $\widetilde{t_0}$ (determined self consistently). Different markers refer to different number $N$ of QPCs employed. For a given $t_0$, increasing $N$ necessitates a larger $t$ (closer to unity) - resulting in smaller charge $q_0$. Therefore, the curve $q_0(\widetilde{t_0})$ climbs toward $e$ with decreasing $t_0$, but in a slower fashion for large $N$. This enables the generation of a very *dilute beam of quasiparticles* with charge $e/3$. Experimental study of the interaction of such a *sparse* beam of quasiparticles with a high potential barrier (a pinched QPC) is described elsewhere [11].

The issue of edge state equilibration is important in general but especially here since it may affect the analysis of the data. In general, partitioning a full Fermi sea, having chemical potential $\mu$, by a QPC with energy dependent transmission $t(\varepsilon)$, results with a non-equilibrium distribution of the transmitted particles. This non-equilibrium distribution may, however, relax (via particle-particle or particle-phonon scattering) into a new sea, characterized by a new chemical potential:

$$\bar{\mu} = \mu \cdot \frac{1}{\Delta \varepsilon} \int t(\varepsilon) d\varepsilon \equiv \mu \bar{t} \ . \qquad (7)$$

Here $\bar{t}$ represents the average (static) transmission, namely, the total fraction of quasiparticles transmitted in the energy 'transport window' $\Delta \varepsilon = qV$, with $V$ being the applied voltage. As long as frequency components of the noise are not 'lost' (by means of capacitive coupling to dissipative environment), the original fluctuations in state occupation are transformed into fluctuations of $\tilde{\mu}$, and noise measurements in a single QPC are not sensitive to equilibration. However, an already partially filled and non-relaxed Fermi sea that impinges on a QPC, at a finite temperature and for energy dependent transmission, results with different transmission and noise compared to that of a fully relaxed sea.

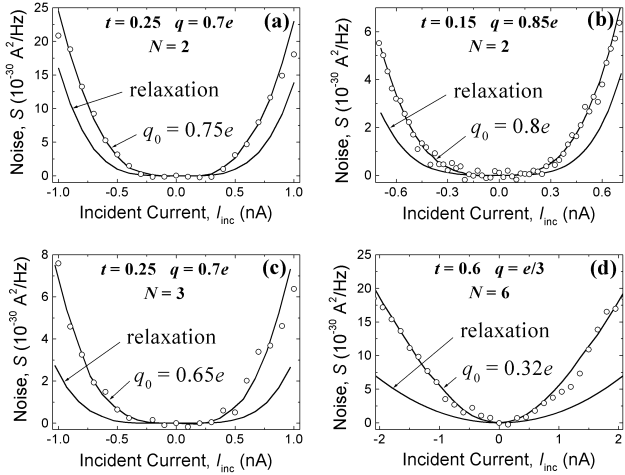

FIG. 4. Comparison of various noise measurements with a model taking into account intra-edge relaxation of the non-equilibrated distribution generated by scattering. The $t$ and $q$ in each panel are of an individual QPC, and $N$ is the number of QPCs in the array. Evidently, the expected noise after relaxation is much smaller than the measured noise.

Since the voltage $V_j$ in the expression for the noise represents the highest energy of an occupied state that approaches the $j^{\text{th}}$ QPC, it equals $V$ for all $j$ without relaxation, but drops between two consecutive QPCs in case of relaxation - with different transmission and noise. Measuring the nonlinear transmission $t_j(\varepsilon)$ and the charge $q_j(t_j)$ of each QPC, the total noise of the equilibrated system is calculated without fitting parameters.

Several examples of noise measurements are shown in Fig. 4 together with the predicted noise when relaxation does take place. We choose distinct cases where the applied voltage $V$ is large and the overall transmission $t_0$ is small, to make the effect of relaxation pronounced. The measured noise deviates clearly from the predicted one for an equilibrated beam and leads, as above, to a charge $q_0 = q(t)$. This behavior does not depend on the distance between the QPCs or on the presence of gates in the path of the edge states. We thus believe that an insignificance intra-edge relaxation occurs over macroscopic distances of hundreds of microns. Other experiments aiming at measuring relaxation were focused on *inter-edge* relaxation, and found relaxation distances between different Integer QH states of hundreds of microns [12], while they were only a few microns long in the FQH regime [13]. We are not aware of measurements of *intra-edge* equilibration length.

In summary, employing shot noise measurements in a Fractional Quantum Hall system, partitioned by an array of identical scattering centers, we find that *bunching* of quasiparticles is determined by the local potential at a single scatterer - and *not* by the total scattering strength of the chain. Hence, we find $e/3$ quasiparticles traversing *opaque* barriers made of a few weak backscatterers. We also find intra-edge equilibration length of a partitioned $e/3$ edge to exceed hundreds of microns.

*Acknowledgements* The work was partly supported by the Israeli Academy of Science and by the German-Israel Foundation (GIF) grant. We thank D. Mahalu for the e-beam lithography.